\documentstyle[12pt]{article}

\advance \textheight by \topskip
\makeatletter
\@addtoreset{equation}{section}
  
\makeatother

\def\reff #1{(\ref{#1})}
\def\({\left(}
\def\){\right)}
\def\beq{\begin{equation}}
\def\eeq{\end{equation}}
\def\={\approx0}

\begin{document}

\title{\bf Zitterbewegung and reduction: 
4D spinning particles and 3D anyons on 
light-like curves}

\author{
Sergei Klishevich${}^a$\thanks{E-mail: klishevich@mx.ihep.su}, 
Mikhail Plyushchay${}^{b,a}$\thanks{E-mail: mplyushc@lauca.usach.cl}\\
{\small ${}^{a}${\it Institute for High Energy Physics, Protvino,
Moscow region, 142284 Russia}}\\
{\small ${}^{b}${\it Departamento de F\'{\i}sica, 
Universidad de Santiago de Chile,
Casilla 307, Santiago 2, Chile}}}
\date{}

\maketitle
\vskip-1.0cm

\begin{abstract}
We construct the model with light-like world-lines
for the massive $4D$ spinning particles and $3D$ anyons.  
It is obtained via the formal
bosonization of pseudoclassical model for the massive Dirac
particle with subsequent reduction to the light-like curves.  
The peculiarity of the light-like trajectories  
produced due to the Zitterbewegung is explained from the viewpoint of
reduction and reparametrization invariance.
\vskip2mm
\noindent 
{\it PACS number(s):}  11.10.Kk, 11.10Ef, 03.20.+i
\vskip2mm
\noindent
{\it Keywords:} Zitterbewegung; Reduction; Spinning particles; Anyons
\end{abstract}

\newpage

\section{Introduction}
The quantum Zitterbewegung plays important but hidden role in
physics of spin particles.  
It reveals itself nontrivially, e.g., 
under description of interaction of Dirac
particle with electromagnetic field 
at the relativistic quantum mechanical level. 
With minimal coupling prescription, the electromagnetic
field interacts locally with the point charge whose coordinates
exhibit the Zitterbewegung \cite{bb},
whereas the attempt of
reformulating the interaction in terms of Pryce-Newton-Wigner
coordinates \cite{pnw} having no trembling component in their evolution
makes the theory to be nonlocal \cite{fw}. 

The superposition of rectilinear and trembling
motions may result in helical 
trajectory for the quantum Dirac particle \cite{hol}. 
It is this picture of motion that is a
characteristic feature of various classical models 
of $4D$ spinning particles \cite{bar}--\cite{tom}
and
$3D$ anyons \cite{any}--\cite{cp2}.  
However, these models with the underlying Zitterbewegung 
suffer from the same problem which emerges for
the pseudoclassical models of $4D$ 
higher spin particles \cite{hi}.
They, unlike the simplest pseudoclassical model for Dirac particle 
\cite{pseudo},
fail to be consistent under attempt to switch on interaction
with an arbitrary gauge or gravitational field.

The success of spin-$1/2$ pseudoclassical 
model \cite{pseudo} indicates that more detailed analysis of the 
hidden role of the Zitterbewegung in the theory of spinning particles
may be useful. In model \cite{pseudo}, 
the velocity of the particle contains a term
being a classical analog of the quantum Zitterbewegung.
Such term has a pure nilpotent nature (see the next Section)
and, as a consequence, generally the motion of the particle
in model \cite{pseudo} cannot be visualized.  
Besides, 
the direct classical analog of the Dirac Hamiltonian
$H=\gamma^0(p_i\gamma_i+m)$ generating the Zitterbewegung 
is given there by a pure nilpotent quantity.
The difficulties with nilpotency
are absent in the models with even spin variables.
It was observed \cite{zit} that classically the velocity
of the massive spinning particle in such models may take any value
up to the velocity of light. With superluminal velocities
for the massive particle, unlike the massless case \cite{m0},
one looses the time interpretation 
for $x^0$ component of the vector $x^\mu$ \cite{zit}. 
It was also shown \cite{zit,m0} that the classical
Zitterbewegung is not gauge-invariant and the modified
coordinates of the particle revealing no trembling motion
in their evolution are the classical analogs of the 
non-covariant Newton-Wigner coordinates. 
But in Dirac theory 
the usual (covariant) coordinates revealing the Zitterbewegung
play more fundamental role from the viewpoint of interactions.
Therefore, the classical models for spinning particles
which would reproduce exactly the peculiar
nature of the quantum Zitterbewegung ---
the instant velocity of the massive particle 
equal to the velocity of light ---
could play a special role in the theory.

Recently, the geometrical model with higher
derivatives has been constructed in ref. \cite{nr}.  
Unlike the 
known geometrical models of $4D$ spinning particles and $3D$
anyons, it is formulated from the very beginning on
the light-like curves.  
Like the known models with higher derivatives \cite{maj}, 
the model \cite{nr} has a spectrum similar to that of the Majorana equation, 
which possesses the massive, massless and tachyonic solutions.  
After reduction by extra constraint singling out a massive state,
one can arrive finally at the
model of massive spinning particle with light-like world trajectories.

In this letter we shall show that the model for massive $4D$
spinning particles and $3D$ anyons with light-like world-lines
may be constructed in another way: by reducing the model of
spinning particles of a fixed mass to the light-like curves. The
model obtained in this way does not contain higher derivatives
and has no hidden tachyonic states which could reveal themselves
after switching on interaction. It has a property of universality
which manifests itself in  the same form for the Lagrangian 
in $4$ and $3$ dimensions.
Finally, it allows us to explain the peculiarity of the light-like 
trajectories produced due to the Zitterbewegung from the viewpoint of
reduction and reparametrization invariance.

The paper is organized as follows.  In Section 2 we shortly
discuss the Zitterbewegung in pseudoclassical model \cite{pseudo} and
show how the model of spinning particle \cite{zit} can be obtained via the
formal bosonization of the former model.  In Section 3 we obtain
the massive model of spinning particle with light-like world
trajectories by the reduction to the light-like curves.  
We discuss different reduction procedures giving rise
to the same final Lagrangian of a simple form which explicitly
guarantees the velocity of light for the particle.  Section 4
concerns the quantum theory of the model in $4$ and $3$ dimensions.
Section 5 is devoted to discussion and concluding remarks.

\section{Bosonization and Zitterbewegung}

The pseudoclassical model for the massive Dirac particle
in $4$ dimensions is given by the Lagrangian \cite{pseudo}
\beq\label{d}
L=\frac{1}{2e}(\dot{x}_\mu-i\lambda \xi_\mu)^2-
\frac{e}{2}m^2-im\lambda\xi_5-\frac{i}{2}\xi_\mu\dot{\xi}{}^\mu
-\frac{i}{2}\xi_5\dot{\xi}_5,
\eeq
where $\xi_\mu$ and $\xi_5$ are Grassmann (odd)
variables, which are  transformed after quantization 
(up to the numerical factors) into $\gamma_5\gamma_\mu$
and $\gamma_5$ matrices.
Variables $e$ and $\lambda$ are even and odd Lagrange multipliers
associated with reparametrization symmetry 
and local supersymmetry of the system
generated by the first-class constraints
\beq\label{d1}
p^2+m^2\=,\quad
p\xi+m\xi_5\=
\eeq
by means of the Poisson-Dirac brackets $\{x_\mu,p_\nu\}=\eta_{\mu\nu}$,
$\{\xi_\mu,\xi_\nu\}=-i\eta_{\mu\nu}$,
$\{\xi_5,\xi_5\}=-i$.
Corresponding Hamiltonian equations of motion for 
the space-time coordinates of the particle
have the form 
\beq\label{dz}
\dot{x}_\mu=ep_\mu+i\lambda\xi_\mu,
\eeq
i.e. in addition to the translational part along the 
energy-momentum vector $p_\mu$,
the velocity has additional term being classical
analog of the quantum Zitterbewegung.
However, in contrast with the quantum case,
this additional term has a nilpotent nature due to which
the classical Zitterbewegung cannot be characterized numerically
and in general the motion of the particle cannot be visualized.

To visualize the classical Zitterbewegung,
let us realize the formal bosonization
of the pseudoclassical model by substituting
the odd Grassmann (classical fermionic) 
variables $\xi_\mu$, $\xi_5$ and $\lambda$
for the even (classical bosonic) variables $q_\mu$, $q_5$ and $v$.
With such substitution the kinetic term
for spin variables $q_\mu$ and $q_5$
will be a total derivative. 
To find the appropriate kinetic term,
it is necessary to note that for
the Grassmann variables there is no usual concept of the length.
The bosonized elongated velocity 
$\tilde{\dot{x}}_\mu=\dot{x}_\mu-vq_\mu$ 
does not feel the length of the vector $q_\mu$ either since
the local change, $q_\mu\rightarrow \gamma q_\mu$,
$\gamma=\gamma(\tau)\neq 0$,
can be compensated by the transformation $v\rightarrow
\gamma^{-1}v$ leaving $\tilde{\dot{x}}_\mu$ to be invariant.
The term $mvq_5$ is invariant under rescaling of $q_5$
in the same way.
Therefore, one can construct the kinetic term which
would be invariant under local rescaling 
$\tilde{q}_M\rightarrow\gamma \tilde{q}_M$,
where we have introduced the notation $\tilde{q}_M=(q_\mu,q_5)$.
Taking also into account that kinetic term for spin variables
in (\ref{d}) is of the first order in parameter evolution derivative
(and, as a consequence, the corresponding term of the action
is reparametrization invariant),
we take 
$-\alpha\sqrt{\dot{\tilde{n}}{}^2}$
as a kinetic term for the bosonic spin variables, where
$\alpha$ is a nonzero parameter,
$\tilde{n}_M=q^M/\sqrt{\tilde{q}{}^2}$
and we suppose that 
$\tilde{q}{}^2=\tilde{q}{}^M\tilde{q}{}^N
\eta_{MN}>0$, $\dot{\tilde n}{}^2\geq0$, 
$\eta_{MN}=(-1,1,1,1,1)$.
Finally, we arrive at the following Lagrangian for $4D$ spinning particle
\cite{zit}:
\beq\label{l0}
L=\frac{1}{2e}\(\dot x_\mu -vq_\mu\)^2 - \frac{e}2m^2 -
mvq_5-\alpha\sqrt{\dot{\tilde n}{}^2}.
\eeq
At the Hamiltonian level the system (\ref{l0})
is described by the set of first class
constraints
\begin{eqnarray}
\label{primary0}&
\phi_1=\tilde\pi\tilde q\=,\quad
\phi_2=\tilde\pi^2 \tilde q^2 - \alpha^2\=,\quad
\phi_3=p_e\=,\quad
\phi_4=p_v\=,\quad
&\\&\label{secondary0}
\phi_5=p^2 + m^2\=,\quad
\phi_6=pq + m q_5,\quad
\phi_7=p\pi + m \pi_5\=,
&\end{eqnarray}
among which constraints \reff{primary0} are primary and
\reff{secondary0}
are secondary constraints, and the quantity
\beq
H = \frac{e}2 \phi_5 + v \phi_6 + \sum_{a=1}^4 w_a\phi_a
\label{ham}
\eeq
plays the role of the total Hamiltonian with $w_a=w_a(\tau)$,
$a=1,\dots,4,$  being arbitrary functions.
Here $\tilde{\pi}_M$ are the momenta canonically conjugate to
$\tilde{q}_M$, $\{\tilde{q}_M,\tilde{\pi}_N\}=\eta_{MN}$.
What is the physical sense of the nontrivial constraints 
$\phi_1$, $\phi_2$, $\phi_6$ and $\phi_7$?
The last two constraints play the same role as the odd constraint
from (\ref{d1}): effectively they are the `square root' constraints 
from the mass shell constraint $\phi_5$,
$\{\phi_6,\phi_7\}=\phi_5$, and
kill the nonphysical degrees of
freedom associated with the time components $q_0$ and $\pi_0$
whose quantum analogs could produce negative norm states.
Constraint $\phi_1$ generates the above mentioned 
local rescaling transformations and guarantees 
that there are no oscillation radial-like
degrees of freedom which could be associated
with the change of the scale of internal variables.
Therefore, the system can have only the internal
rotational degrees of freedom.
On the surface of other constraints, the constraint $\phi_2$ 
can be written in the equivalent form 
\beq\label{spin}
S^2 -\alpha^2\=,
\eeq
where
$$
S_\mu=\frac1{2\sqrt{-p^2}}\epsilon_{\mu\nu\rho\sigma}p^\nu M^{\rho\sigma}
$$
is the spin vector 
and $M^{\mu\nu}$ is the conserved angular momentum of the
particle. So, like the initial model (\ref{d}),
the bosonized model (\ref{l0}) also describes the particle
with fixed mass and spin.
One notes that the constraint analogous to 
(\ref{spin}) appears in the form of spirality-fixing condition
in the twistor theory  
of massless spin particles \cite{t1}-\cite{t5}.
  
Hamiltonian (\ref{ham}) generates the 
equations of motion
\beq\label{bz}
\dot{x}_\mu=ep_\mu+vq_\mu
\eeq
being the bosonized version of Eq. (\ref{dz}),
i.e. now the classical analog of the quantum Zitterbewegung
can be visualized though it is not gauge-invariant \cite{zit}.
Taking into account constraints $\phi_5$ and $\phi_6$,
Eq. (\ref{bz}) gives $\dot{x}{}^2=e^2m^2+2emvq_5-q^2v^2$.
{}As we have mentioned, the analysis shows that
the velocity of the particle
may take any value up to the velocity of light ($\dot{x}{}^2\leq0$),
but cannot be superluminal ($\dot{x}{}^2>0$) \cite{zit}.

In conclusion of this section let us note
that the obtained model can be transferred
to the $3$-dimensional space-time.
Classically,  the only difference of the $3D$ case
from the $4D$ case consists in the spin nature.
In $3D$ spin has a nature of a pseudoscalar \cite{cp2},
\beq\label{ss}
S=\frac1{2\sqrt{-p^2}}\epsilon_{\mu\nu\rho}p^\mu M^{\nu\rho}.
\eeq
Here the constraint $\phi_2$ is, again,  reduced 
to the condition of the form (\ref{spin})
but with $S$ being pseudoscalar (\ref{ss}).
Thus, all the results of the next Section
on the reduction of the model to the light-like
curves will be valid for both cases of $4$ and $3$ 
dimensions with the described difference which
has to be implied. Essential distinction between
the two cases will happen at the quantum level
and this will be discussed in what follows.

\section{Reduction to the light-like world trajectories}

Let us require that the velocity of the spinning particle
like the instant velocity of the quantum Dirac particle
would be equal to the velocity of light.
This can be achieved at the Lagrangian and Hamiltonian levels
with exactly the same final result.
First, let us impose the condition
\beq\label{chi}
\chi=e^2 m^2 + 2 e m v q_5 - q^2 v^2\=
\eeq
providing $\dot x^2=0$ as a gauge condition.
It can be represented equivalently as 
\beq\label{q5}
q_5-\frac{1}{2}(q^2ve^{-1}m^{-1}-emv^{-1})\=.
\eeq
We can treat this condition as a gauge for the 
constraint $\phi_1$
and exclude the variables $q_5$ and $\pi_5$ from the system
with the help of
the set of second class constraints 
$\phi_1$ and (\ref{q5}) 
by means of the Dirac brackets
or by reducing symplectic two-form of the system,
$\omega=dp_\mu\wedge dx^\mu+d\pi_\mu\wedge dq^\mu
+d\pi_5\wedge dq_5$,
to the surface of these second-class constraints.
Then, performing the inverse Legendre transformation
for the reduced Hamiltonian, 
we arrive at the Lagrangian
\begin{equation}
\label{L1}
L = \frac1{2e}\dot x^2 - \dot xq V
- \frac{2\alpha m}{q^2 V^2+m^2}
\sqrt{\dot q^2 V^2 + 2\dot qq \dot{V}
V + q^2\dot{V}{}^2},
\end{equation}
where $V=v/e$. This Lagrangian can also be derived by the
direct substitution of (\ref{q5}) into Eq. (\ref{l0}).
Obviously, Lagrangian (\ref{L1}) describes light-like
motion of the particle since the equation of motion for $e$
gives directly the necessary relation $\dot{x}{}^2=0$.
At the Hamiltonian level the reduced system (\ref{L1})
is described by the set of first class constraints 
\begin{eqnarray}
&\psi_1=\pi q-p_V V\=,\quad
\psi_2=\pi^2(q^2V^2+m^2)^2-4m^2\alpha^2V^2\=,\quad
\psi_3=p_e\=,&\label{c1}\\
&
\psi_4=2pq V+q^2V^2+p^2\=,\quad
\psi_5=p\pi+p_V V^2\=,\quad
\psi_6=p^2+m^2\=,&\label{c2}
\end{eqnarray}
where constraints (\ref{c1}) are primary and
Eq. (\ref{c2}) gives the set of secondary
constraints. From the explicit form of the constraints
it is clear that the canonical pair
$V$ and $p_V$ effectively plays the role
of the removed scalar variables $q_5$ and $\pi_5$.
In particular, the constraint $\psi_2$ playing  
the role of spin-fixing condition
may be represented in the form (\ref{spin})
if to take into account other constraints.

One can simplify Lagrangian (\ref{L1}) if to note that
it depends on $q_\mu$ and $V$
only via the combination $Vq_\mu$.
Introducing the dimensionless vector
$Q_\mu=Vq_\mu/m$, we represent Lagrangian
(\ref{L1}) in the equivalent form
\beq\label{L2}
L = \frac{\dot x^2}{2e} - m\dot{x}Q
- \frac{2\alpha}{1+Q^2}\sqrt{\dot{Q}{}^2}.
\eeq
At the Hamiltonian level the physical equivalence of the
reduced system (\ref{L2}) to the initial
system (\ref{l0}) may be established in 
the  following way.
In this case 
the complete set of primary and secondary 
Hamiltonian constraints 
is the set of $4$ first class constraints:
\begin{eqnarray}
&\varphi_1=p^2+m^2\=,\quad 
\varphi_2=\Pi^2(Q^2+1)^2-4\alpha^2\=,&\label{C1}\\
&\varphi_3=pQ+\frac{1}{2}m(Q^2-1)\=,\quad
\varphi_4=p\Pi+Q\Pi\=,&
\label{C2}
\end{eqnarray}
where $\Pi_\mu$ are the momenta canonically conjugate
to $Q^\mu$.
On the surface of other constraints,
the constraint $\varphi_2$ is reduced to the
spin-fixing condition (\ref{spin}).
One can introduce
the conditions $\chi_1=\Pi Q\=$ and $\chi_2=Q^2-1\=$
as the gauges for
the constraints $\varphi_3$ and $\varphi_4$.
This set of 4 second class constraints is equivalent
to the set of constraints $pQ\=$,
$p\Pi\=$, $\Pi Q\=$ and $Q^2-1\=$.
One can arrive at the same set of second class constraints
and first class constraints $\varphi_1$ and $\Pi^2-\alpha^2\=$
if to introduce at the Hamiltonian level into the model (\ref{l0})
the gauges $q_5\=$ and $\pi_5\=$.
Indeed, these two gauges form themselves the set
of second class constraints.
Reducing the system to the surface defined by them
we shall get the mixed set of first and second class constraints
specified above with substitution of $Q_\mu$ and $\Pi_\mu$ for
$q_\mu$ and $\pi_\mu$.

One can also arrive at the Lagrangian (\ref{L2}) if
to treat relation (\ref{chi}) as a constraint
on $v$. Then, reducing the system to the surface 
of constraints (\ref{chi}) and $p_v\=$, 
and performing the inverse Legendre transformation,
we arrive at the Lagrangian of the form (\ref{l0}) but 
with $v=Ve$, where $V=V(q^2,q_5)$ is defined by the equation
$
V^2q^2 +2mVq_5-m^2=0.
$
Finally, realizing the substitution $Q_\mu=Vq_\mu/m$,
we get the Lagrangian (\ref{L2}).

Due to the invariance of the terms
$vq_\mu$, $vq_5$ and $\sqrt{\dot{\tilde{n}}{}^2}$
in (\ref{l0})
under the local transformations 
$\tilde{q}_M\rightarrow \gamma \tilde{q}_M$,
$v\rightarrow \gamma^{-1}v$,
one can remove the Lagrange multiplier $v$ by introducing
new variables $\tilde{Q}_M=m^{-1}ve^{-1}\tilde{q}_M$
before the reduction procedure.
Then the Lagrangian (\ref{l0})
can be represented in the physically equivalent form
\beq\label{L3}
L=\frac{\dot{x}{}^2}{2e}-m\dot{x}Q-
\frac{e}{2}m^2(1+2Q_5-Q^2)-\sqrt{\dot{\tilde{N}}{}^2},
\eeq
where $\tilde{N}_M=\tilde{Q}_M/\sqrt{\tilde{Q}{}^2}$.
Now, the reduction to the light-like curves may be realized
by the substitution 
$
Q_5=\frac{1}{2}(Q^2-1)
$
into Lagrangian (\ref{L3}). The result is given by  
the same Lagrangian (\ref{L2}).

We have reduced the model of spinning particle to
the light-like curves defined by the relation
$\dot{x}{}^2=0$. 
Is it possible to use for reduction 
the relativistic-invariant condition
of more general form 
\beq\label{nu}
\dot{x}{}^2=-\nu^2
\eeq
with $\nu$ being a constant?  The answer is `no'
and the special role of the light-like case ($\nu=0$) may be
understood as follows.  If $\nu\neq 0$, then by changing the
scale of the parameter evolution, $\tau\rightarrow\vert\nu\vert^{-1}\tau$, we
change condition (\ref{nu}) for the relation $\dot{x}{}^2=-1$. The
latter relation is nothing else as the proper time gauge
condition, i.e. for $\nu\neq 0$ Eq. (\ref{nu}) destroys
the reparametrization invariance and fixes the parameter evolution.
The special role of $\nu=0$ case in Eq. (\ref{nu}) can also be
understood from the observation that if we change  $\dot{x}{}^2$ in
Lagrangian (\ref{L2})
for $\dot{x}{}^2-\nu^2$,
$\nu\neq 0$, the equation of motion for $e$ will be Eq. (\ref{nu}), but
the term $-\nu^2/2e$ in the Lagrangian will destroy
the reparametrization invariance of the action.  Therefore, we
conclude that introducing relativistic-invariant condition
(\ref{nu}) into the theory for $\nu\neq 0$ is the
parametrization-fixing procedure, whereas reduction with $\nu=0$
is the only possible case to be consistent with
the reparametrization invariance of the theory for the massive spinning
particle.  This observation gives an explanation for the
paradox of the special role played by the light-like
world-lines in the theory of massive spinning particles and sheds
some light on the nature of the Zitterbewegung.

\section{On quantum theory of the model}

We have demonstrated that the realized
reduction procedure giving rise to the 
model of massive spinning particle in $4$ and $3$ dimensions
does not change the physical content
of the initial system. 
Therefore, the results on the quantum theory
of the model (\ref{l0}) in $4$ and $3$
dimensions can be used. The interested reader may found 
the details on the covariant and reduced phase space
quantization of the system (\ref{l0}) for these two cases in refs.
\cite{zit} and \cite{any}, but here we restrict ourselves only
by a short comment on the results 
stressing their topological aspects.

In the case of $4$ dimensions before taking into account 
the spin-fixing condition (\ref{spin}), 
the internal configuration space of the system
can be described effectively by one $3$-dimensional unit vector,
i.e. it is a sphere $S^2$, whereas the internal phase space 
is the cotangent bundle $T^*S^2$. As a consequence,
the spin operator has a nature of orbital angular momentum \cite{pr}.
Then taking into account 
spin-fixing condition (\ref{spin}) leads to the quantization 
of the parameter $\alpha$:
$\alpha=n(n+1)$, $n=1,2,\ldots$.
As a result, spin of the particle 
takes the corresponding integer value $s=n$.

In the case of $3$ dimensions, the internal configuration
space of the system is described effectively by a $2$-dimensional 
unit vector, i.e. it is a sphere $S^1$ (circle).
The internal phase  space 
(before reduction to the  spin-fixing surface (\ref{spin}))
is the cotangent bundle $T^*S^1$, 
and the eigenvalues of the quantum spin operator can take
arbitrary values on the real line \cite{pr}.
No quantization condition
appears for the parameter here.
Due to the quantum analog of the 
spin-fixing condition (\ref{spin}), spin of the particle
takes two values $s=\pm j$ in the case when
$\vert \alpha\vert=j$ and $j>0$ is integer or half-integer
number. For $\alpha \in {\bf R}$,
$\vert\alpha\vert \neq j$,
the violation of the classical $P$-invariance
happens at the quantum level  and spin takes either
the value $s=+\alpha$ or $s=-\alpha$.

To describe in $4D$ half-integer spins,
the system has to be modified by using two
sets of spin variables \cite{top}, $\tilde{q}{}^i_M$, $i=1,2$, instead of
one set used here. In this way the internal configuration space
of the system may be described effectively by the pair of mutually orthogonal
unit vectors, i.e. topologically it will be 
equivalent to the manifold of doubly-connected $SO(3)$ group.
Such a difference is sufficient 
to describe integer and half-integer spin states 
at the quantum level \cite{pr}.
For the appropriately modified  model \cite{top} 
the reduction to the light-like curves 
may be realized in the way similar
to the described here.

It is worth to note that at the quantum level
the quantization of the pseudoclassical model (\ref{d})
in $4D$ and $3D$ does not reveal  so essential
difference as for the obtained bosonic model.
For the system (\ref{d}), the only difference 
is that in $3D$ the model gives rise to 
the pair of Dirac equations 
corresponding to spin $s=+1/2$ and $s=-1/2$ 
instead of one Dirac equation in $4D$ \cite{gp}.

\section{Discussion and outlook}

In the analysis of the classical
analog of the quantum Zitterbewegung we started from the 
(locally) supersymmetric pseudoclassical Lagrangian (\ref{d}) 
describing spin-$1/2$ Dirac particle. 
As a final result, we arrived at the bosonic Lagrangian
(\ref{L2}), which universally describes $4D$ spinning particles 
and $3D$ anyons on the light-like curves.
In both cases spin is defined by
the parameter $\alpha$ which, due to the topological reasons,
is quantized in $4D$ but 
can take arbitrary real values in $3D$.

Constructing the intermediate bosonic Lagrangian (\ref{l0}),
we essentially exploited the 
properties of (locally) supersymmetric spin-1/2 particle action. 
In other words, our derivation  was based on (\ref{d})
not only as on a starting point, 
but the applied bosonization 
procedure itself was governed by the nature of the initial 
spin-1/2 particle supersymmetric action.
With the help of the final form of the Lagrangian (\ref{L2}),
obtained via reduction of (\ref{l0}),
we observed that the peculiarity
of the light-like trajectories produced due to the
Zitterbewegung can be explained from the viewpoint of 
interplay of reparametrization invariance and reduction.

In our constructions we did not insist 
on attaching physical meaning to the Grassmann variables
since one is only supposed to compute physical quantities
after quantization. The impossibility to describe the Zitterbewegung
in pseudo-classical approach was used here as one of the motivations
for its analysis with which we found that
the special role of the light-like trajectories for the massive 
spinning particles can be understood at the classical level.

To conclude, let us indicate some problems that deserve
further attention.

First, we note that a priori there is no obstruction
for applying the ideas of the formal bosonization constructions
to the case of locally supersymmetric string. 
What will be the physical content of the bosonic string 
constructed in such a way?

The spin-fixing constraint (\ref{spin}) is reminiscent of the
spirality-fixing condition in twistor formulation of
the massless spin particles \cite{t1}-\cite{t5}. 
It would be interesting to investigate 
the Zitterbewegung aspect for the massive spinning 
particles using the twistor approach. However, it is clear 
that the $4D$-$3D$ universality of our model (\ref{L2}) 
would be not so explicit in the twistor theory, 
in which two types of spinors 
have to be distinguished in 4D, 
whereas the 3D case is characterized by the 
presence of only one type of spinors.

The reduction procedure used for the system (\ref{l0}) 
cannot be applied directly for the $P$-noninvariant models 
of $3D$ anyons \cite{f,cp2}. The reason is that their corresponding
Lagrangians do not contain the analog of scalar variable $q_5$. 
Probably, to reduce those models to the light-like curves, 
they first should be modified appropriately by introducing
some auxiliary scalar variable.
By the same reason it is not clear either how the
corresponding reduction procedure could be realized
within the framework of the twistor approach applied
for describing massive spin particles.

Finally, it seems to be interesting to continue the investigation
of the model given by the Lagrangian with light-like
world-lines (\ref{L2}) in the context of switching on
interactions.

\vskip0.5cm
{\bf Acknowledgment}
\vskip3mm

M.P. has been supported in part
by the grant 1980619 from FONDECYT (Chile)
and by DICYT (USACH).


\begin{thebibliography}{**}

\bibitem{bb}
A. O. Barut and A. J. Bracken,
{Phys. Rev.} { D 23} (1981) 2454.

\bibitem{pnw}
M. H. L. Pryce, { Proc. Roy. Soc.}
{ 150 A} (1935) 166, { 195 A} (1948) 62;\\
T. D. Newton and E. P. Wigner,
{ Rev. Mod. Phys.} { 21} (1949) 400.

\bibitem{fw}
L. L. Foldy and S. A. Wouthuysen, { Phys. Rev.}
{ 78} (1950) 29.

\bibitem{hol}
P. R. Holland, {\it The Quantum Theory of Motion:
An Account of the de Broglie-Bohm Causal Interpretation
of Quantum Mechanics} (Cambridge Univ. Press, 1995).

\bibitem{bar}
A. O. Barut and N. Zanghi,
{ Phys. Rev. Lett.} { 52} (1984) 2009;\\
A. O. Barut and M. Pavsic,
{ Phys. Lett.} { B 216} (1987) 297.

\bibitem{m0}
M. S. Plyushchay, { Mod. Phys. Lett.}
{ A 4} (1989) 837;
{ Phys. Lett.} { B 243} (1990) 383.

\bibitem{zit}
M. S. Plyushchay, { Phys. Lett.} 
{ B 236} (1990) 291.

\bibitem{top}
M. S. Plyushchay, { Phys. Lett.}
{ B 248} (1990) 299.

\bibitem{pav}
M. Pavsic, E. Recami and W. A. Rodriguez, Jr.,
{ Hadronic J.} { 18} (1995) 98,
{quant-ph/9803036}.

\bibitem{tom}
S. L. Lyakhovich, A. Yu. Segal and  A. A. Sharapov, 
{ Phys.Rev.} { D 54} (1996) 5223, hep-th/9603174. 

\bibitem{any}
M. S. Plyushchay, { Phys. Lett.} 
{ B 248} (1990) 107.

\bibitem{maj}
M. S. Plyushchay, { Phys. Lett.} 
{ B 262} (1991) 71;
{ Nucl. Phys.} { B 362} (1991) 54;\\
Yu. A. Kuznetsov and M. S. Plyushchay,
{ Nucl. Phys.} { B 389} (1993) 181. 

\bibitem{f}
M. S. Plyushchay,
{ Int. J. Mod. Phys.}
{ A 7} (1992) 7045.


\bibitem{jn}
R. Jackiw and V. P. Nair,
{ Phys. Rev. Lett.} { 73} (1994) 2007,
{hep-th/9403010}.

\bibitem{cp2}
J. L. Cort\'es and M. S. Plyushchay, { Int. J. Mod. Phys.} 
{ A 11} (1996) 3331, {hep-th/9505117}.

\bibitem{hi}
P. P. Srivastava, { Nuovo Cimento Lett.} { 19}
(1977) 239;\\
V. D. Gershun and V. I. Tkach,
{ JETP Lett.} { 29} (1979) 320;\\
A. Barducci and L. Lusanna, { J. Phys.}
{ A 16} (1983) 1993;\\
P. S. Howe, S. Penati, M. Pernici
and P. Townsend,
{ Phys. Lett.} { B 215} (1988) 555.

\bibitem{pseudo}
F. A. Berezin and M. S. Marinov,
{ JETP Lett.} { 21} (1975) 678;
{ Ann. Phys.} { 104} (1977) 336;\\
L. Brink, S. Deser, B. Zumino, P. DiVecchia and P. Howe,
{ Phys. Lett.} { B 64} (1976) 435;\\
L. Brink, P. DiVecchia and P. S. Howe,
{ Nucl. Phys.} { B 118} (1977) 76.

\bibitem{nr}
A. Nersessian and E. Ramos,
{ Phys. Lett.} { B 445} (1998) 123, hep-th/9807143;
hep-th/9812077.

\bibitem{t1}
D. P. Sorokin, V. I. Tkach and D. V. Volkov,
Mod. Phys. Lett. A3 (1989) 901;\\
D. P. Sorokin, V. I. Tkach, D. V. Volkov and A. A. Zheltukhin,
Phys. Lett. B216 (1989) 302.

\bibitem{t2}
Y. Eisenberg and S. Solomon, Nucl. Phys. B309 (1988) 709;
Phys. Lett. B220 (1989) 562.

\bibitem{t3}
M. S. Plyushchay, Phys. Lett. B240 (1990) 133.

\bibitem{t4}
A. S. Galperin, P. S. Howe and K. S. Stelle,
Nucl. Phys. B368 (1992) 248, hep-th/9201020.

\bibitem{t5}
P. S. Howe and P. C. West,
Int. J. Mod. Phys. A7 (1992) 6639.

\bibitem{pr}
M. S. Plyushchay and A. V. Razumov,
{ Int. J. Mod. Phys.} { A 11} (1996)
1427, hep-th/9306017. 

\bibitem{gp}
J. Gamboa and M. Plyushchay,
{ Nucl. Phys.} { B 512}
(1998) 485, hep-th/9711170.



\end{thebibliography}
\end{document}